\documentclass[conference]{IEEEtran}
\ifCLASSINFOpdf
\else
\usepackage{graphicx}
\graphicspath{{../eps/}}
\DeclareGraphicsExtensions{.eps}
\fi

\usepackage{amsmath}
\usepackage{amsfonts,amssymb}
\usepackage{amssymb}
\usepackage{wrapfig}
\usepackage{psfrag}
\usepackage{epstopdf}
\usepackage{cite}
\usepackage{graphicx}
\usepackage{subfigure}
\usepackage{threeparttable}
\usepackage{cases}
\usepackage{subeqnarray}
\usepackage{color}
\usepackage{underscore}
\usepackage{verbatim}
\usepackage{bm}
\usepackage{stfloats}
\usepackage{caption}

\usepackage{algorithm}
\usepackage{algorithmic}

\begin{document}
	
	\title{Environment-Aware Beam Selection for IRS-Aided Communication with Channel Knowledge Map}
	%
	%
	%
	
	\author{\IEEEauthorblockN{$\text{Dingyang~Ding}^*$, $\text{Di~Wu}^*$, $\text{Yong~Zeng}^*\ddagger$, $\text{Shi~Jin}^*$, and $\text{Rui~Zhang}^\dagger$}\\
		\IEEEauthorblockA{*National Mobile Communications Research Laboratory, Southeast University, Nanjing 210096, China\\
			              $\ddagger$Purple Mountain Laboratories, Nanjing 211111, China\\
                          $\dagger$Department of Electrical and Computer Engineering, National University of Singapore, Singapore 117583\\
			              220200693@seu.edu.cn, 220200691@seu.edu.cn, yong\_zeng@seu.edu.cn, jinshi@seu.edu.cn, elezhang@nus.edu.sg}
    }
	
	\maketitle

\begin{abstract}
Intelligent reflecting surface (IRS)-aided communication is a promising technology for beyond 5G (B5G) systems, to reconfigure the radio environment proactively. However, IRS-aided communication in practice requires efficient channel estimation or passive beam training, whose overhead and complexity increase drastically with the number of reflecting elements/beam directions. To tackle this challenge, we propose in this paper a novel environment-aware joint active and passive beam selection scheme for IRS-aided wireless communication, based on the new concept of \emph{channel knowledge map} (CKM). Specifically, by utilizing both the location information of the user equipment (UE), which is readily available in contemporary wireless systems with ever-increasing accuracy, and the environment information offered by CKM, the proposed scheme achieves efficient beam selection with either no real-time training required (\emph{training-free beam selection}) or only moderate training overhead (\emph{light-training beam selection}). Numerical results based on practical channels obtained using commercial ray tracing software are presented, which demonstrate the superior performance of the proposed scheme over various benchmark schemes.

\end{abstract}


\IEEEpeerreviewmaketitle
\section{Introduction}
Intelligent reflecting surface (IRS) utilizes a large number of passive reflecting elements to achieve controllable amplitude and/or phase of reflected signals and thereby reconfigurable wireless channels. As such, IRS-aided wireless communication has received significant attention recently and been thoroughly investigated in the literature  \cite{1,2,3,4,5}. Thanks to its low hardware and energy cost, intrinsic full-duplex operation, and compatibility with existing wireless systems, IRS has been regarded as a promising technology to enhance the capacity for beyond 5G (B5G) communication systems.\par
However, to practically realize the promising benefits of IRS-aided communication, efficient channel state information (CSI) acquisition is required for joint active and passive beamforming design. This is particularly challenging for IRS-aided systems for two reasons. On one hand, IRS is a passive device that has neither signal transmission nor reception capability, which makes it infeasible to directly estimate channels with intended base stations (BSs) or user equipments (UEs). On the other hand, IRS typically consists of a very large number of elements that have different channel coefficients, which incurs prohibitive overhead for channel estimation. Recently, extensive research efforts have been devoted to devising efficient schemes for IRS channel estimation. Specifically, existing schemes can be roughly classified into two categories, with dedicated sensing devices installed on IRS \cite{6,7} or not \cite{8,9,10,11,12,13,14,15}. In particular, for the case without any sensing devices at IRS, efficient methods have been proposed to estimate the cascaded channels between the UEs and BS via IRS \cite{8,9,10,11,12}. For example, in \cite{9}, by turning on only one IRS element at each time, the complete cascaded channel is  obtained by successively estimating that associated with each IRS element. However, the CSI estimation accuracy of this method is vulnerable to noise since the received training signal is extremely weak with just one element turned on. In \cite{10}, a subgroup-based channel estimation method was proposed for IRS-aided broadband system. By exploiting the users' channel correlation due to their common BS-IRS channel, a multi-user channel estimation scheme was proposed in \cite{11}. In \cite{12}, by assuming that the BS-IRS channel varies at a much larger time scale than the IRS-UE channel, a two-time-scale channel estimation scheme was proposed. Moreover, in \cite{13,14,15}, various compressive sensing-based cascaded channel estimation schemes were proposed by exploiting the BS-IRS channel spatial sparsity or low-rank property.\par
Instead of explicitly estimating the CSI of IRS, another practically appealing approach is to search for the optimal IRS passive beam directly for each UE given a predefined beam codebook \cite{16}. However, the resultant beam training overhead  still increases rapidly with the size of the codebook to ensure high resolution. In this paper, we propose a new approach for efficient beam selection in IRS-aided systems, by exploiting the environment-aware communication enabled by \emph{channel knowledge map} (CKM) \cite{17}. CKM is a site-specific database, which aims to provide location-specific channel information (such as channel gain, shadowing, path angle, delay, etc.) which reflects the actual radio propagation environment where the UE is located. In \cite{18}, environment-aware and training-free beam alignment schemes were proposed for millimeter wave massive multiple input multiple output (MIMO) systems based on two types of CKM, namely \emph{channel path map} (CPM) and \emph{beam index map} (BIM). For IRS-aided communication considered in this paper, we propose a BIM-enabled environment-aware joint active and passive beam selection scheme, which requires either no real-time channel training (\emph{training-free beam selection}) or only limited channel training  (\emph{light-training beam selection}). Specifically, with the proposed training-free beam selection, the beam pair in BIM that was used most frequently by the $K$ neighboring UEs of the target UE is directly selected. In contrast, with the light-training beam selection scheme, an extra beam sweeping procedure is performed over those beam pairs of the $K$ neighbors only, instead of over the complete beam codebook. Simulation results based on practical channels generated by commercial ray tracing software are presented, which show that the proposed BIM-based beam selection scheme significantly outperforms various  benchmark schemes, thanks to its great saving in training overhead with the environment-awareness enabled by BIM.
\section{System Model}
As shown in Fig. 1, we consider an IRS-aided wireless communication system, where an IRS with $N\gg 1$ passive reflecting elements is deployed to assist the communication from the BS to UE that may move in a specific area, denoted as $\mathcal{Q}$. We assume that the BS has $M\gg 1$ antennas and the UE has one antenna. Besides, the direct link between the BS and the UE is assumed to be negligible due to severe blockage. The active and passive beamforming vectors at the BS and the IRS are denoted as $\textbf{f}\in\mathbb{C}^{M\times1}$ and $\textbf{v} = [e^{j\theta_1},...,e^{j\theta_N}]^{T}$, respectively, where $\theta_n\in[0,2\pi)$ is the phase shift of the $n$th element of the IRS. We focus on the codebook-based beamforming, where $\textbf{f}$ and $\textbf{v}$ are selected from pre-determined codebooks $\mathcal{F}$ and $\mathcal{V}$, respectively. The total number of candidate beamforming vectors in $\mathcal{F}$ and $\mathcal{V}$ are respectively denoted as $|\mathcal{F}|$ and $|\mathcal{V}|$.\par
\begin{figure}[h]
\centering
\includegraphics[width=8.6cm,height=4.5cm]{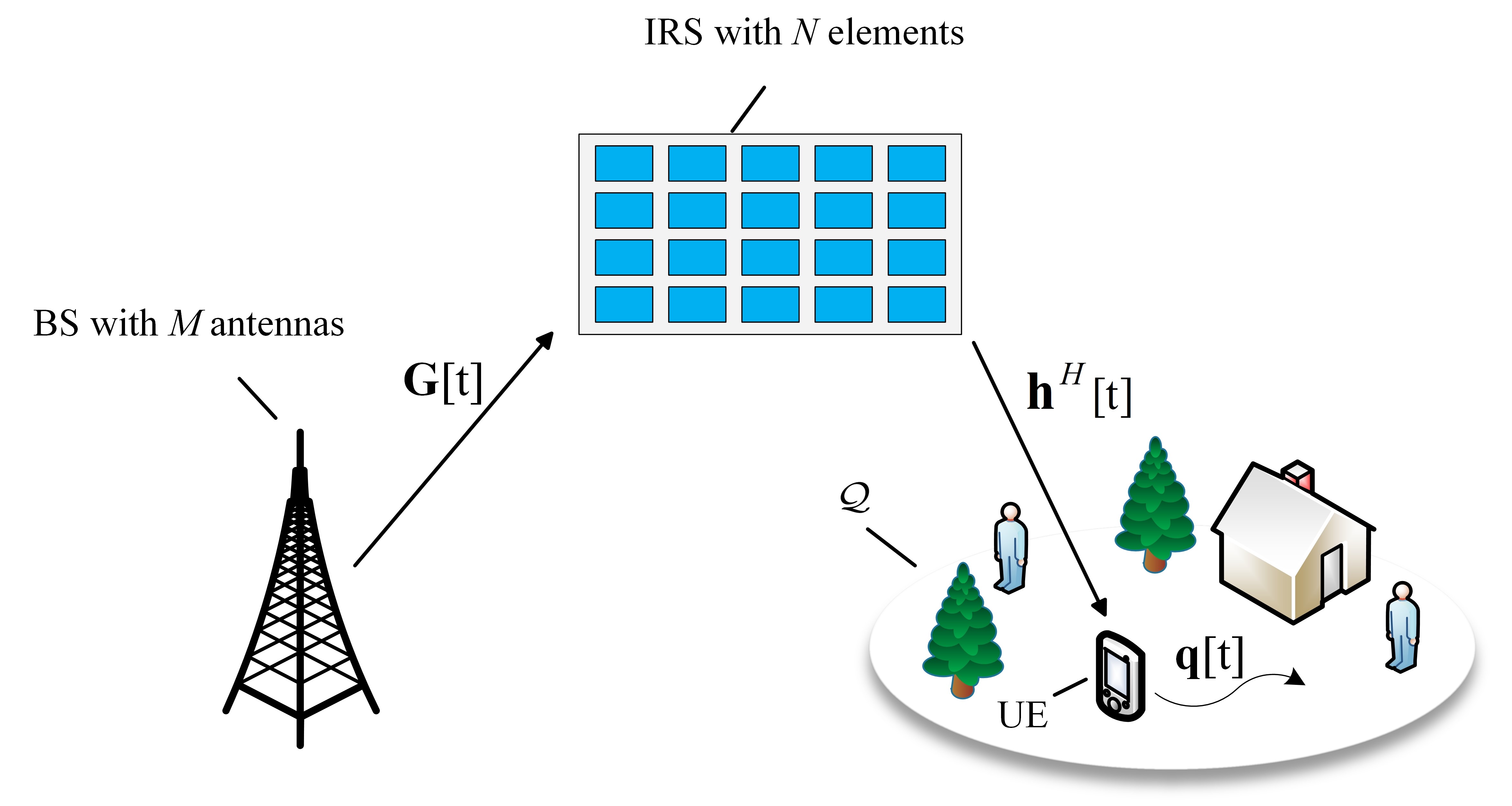}
\caption{Environment-aware beamforming for IRS-aided communication.}
\end{figure}
We assume quasi-static block fading channels, where the BS-IRS and IRS-UE channels remain constant within each channel coherent block that consists of $S$ symbol durations. Denote by $\textbf{G}[t]\in\mathbb{C}^{N\times M}$ and $\textbf{h}^{H}[t]\in\mathbb{C}^{1\times N}$ the BS-IRS and IRS-UE channel coefficients for the channel coherent block $t$, respectively. We focus on downlink communication, where the received signals at the UE over $T$ channel coherent blocks can be written as
 \begin{equation}
   \begin{split}
   y[t,s] = \sqrt{P}\textbf{h}^{H}[t]\bm{\Theta}[t]\textbf{G}[t]\textbf{f}[t]x[t,s] + z[t,s],\\
   s = 1,2,...,S,\quad t = 1,2,...,T,
   \end{split}
 \end{equation}
 where $P$ is the transmit power of the BS, $x[t,s]$ denotes the $s$th information-bearing symbol during channel coherent block $t$ with normalized power, i.e., $\mathbb{E}[|x[t,s]|^2] = 1$, $\bm{\Theta}[t]$ = diag($\textbf{v}[t]$) denotes the reflection coefficient matrix of the IRS, and $z[t,s]\sim\mathcal{CN}(0,\sigma^2)$ denotes the additive white Gaussian noise (AWGN). Besides, $\textbf{f}[t]\in\mathcal{F}$ and $\textbf{v}[t]\in\mathcal{V}$ denote the selected active and passive beamforming vectors during channel coherent block $t$, respectively. By defining $\bm{\Phi}[t]$ = diag($\textbf{h}[t]$)$\textbf{G}[t]\in\mathbb{C}^{N\times M}$ as the equivalent cascaded channel, (1) can be equivalently written as
 \begin{equation}
   \begin{split}
   y[t,s] = \sqrt{P}\textbf{v}^{H}[t]\bm{\Phi}[t]\textbf{f}[t]x[t,s] + z[t,s],\\
   s = 1,2,...,S,\quad t = 1,2,...,T.
   \end{split}
 \end{equation}
 Under the ideal scenario when the perfect CSI of $\textbf{G}[t]$ and $\textbf{h}[t]$ (and hence $\bm{\Phi}[t]$) over all the $T$ channel coherent blocks is available and $S$ is sufficiently large, the maximum average communication rate in bits per second per hertz (bps/Hz) is
 \begin{equation}
   R^* = \frac{1}{T}\sum_{t=1}^{T}\log_2(1 + \bar{P}|\textbf{v}_{\mathrm{opt}}^{H}[t]\bm{\Phi}[t]\textbf{f}_{\mathrm{opt}}[t]|^{2}),
 \end{equation}
where $\bar{P}=P/{\sigma}^{2}$ is the transmit SNR and ($\textbf{v}_{\mathrm{opt}}[t],\textbf{f}_{\mathrm{opt}}[t]$) is the optimal beam pair selected from the codebooks $\mathcal{V}$ and $\mathcal{F}$ for channel coherent block $t$, i.e.,
\begin{equation}
   (\textbf{v}_{\mathrm{opt}}[t],\textbf{f}_{\mathrm{opt}}[t]) = \arg\mathop{\max}_{\textbf{v}\in\mathcal{V},\textbf{f}\in\mathcal{F}}|\textbf{v}^{H}\bm{\Phi}[t]\textbf{f}|^{2}.
 \end{equation}\par
 However, attaining the maximum communication rate in (3) requires perfect CSI, which is quite challenging to acquire for IRS-aided communication in practice. Although various channel estimation methods have been proposed \cite{6,7,8,9,10,11,12,13,14,15}, they typically require high training overhead that increases with the number of IRS elements $N\gg 1$ for each channel coherent block $t$. Denote by $S_{\mathrm{tr}}\leq S$ the number of symbol durations used for channel training at each coherent block. Since the cascaded channel matrix $\bm{\Phi}[t]$ has $MN$ unknown entries to be estimated, we generally require $S_{\mathrm{tr}}\geq MN\gg 1$. Let $\widehat{\bm{\Phi}}[t]$ denote the estimated cascaded channel matrix at channel coherent block $t$, based on which the beam pair ($\hat{\textbf{v}}[t],\hat{\textbf{f}}[t]$) is selected, i.e.,
 \begin{equation}
   (\hat{\textbf{v}}[t],\hat{\textbf{f}}[t]) = \arg\mathop{\max}_{\textbf{v}\in\mathcal{V},\textbf{f}\in\mathcal{F}}|\textbf{v}^{H}\widehat{\bm{\Phi}}[t]\textbf{f}|^{2}.
 \end{equation}\par
 Therefore, with channel training-based beam selection, the effective average communication rate is
 \begin{equation}
 \begin{split}
   R_{\mathrm{tr}} &= \frac{1}{T}\sum\limits_{t=1}^{T}\frac{S-S_{\mathrm{tr}}}{S}\log_{2}(1 + \bar{P}|\hat{\textbf{v}}^{H}[t]\bm{\Phi}[t]\hat{\textbf{f}}[t]|^{2})\\
   &\leq\Big(1 - \frac{MN}{S}\Big)\frac{1}{T}\sum\limits_{t=1}^{T}\log_{2}(1 + \bar{P}|\hat{\textbf{v}}^{H}[t]\bm{\Phi}[t]\hat{\textbf{f}}[t]|^{2})\\
   &\leq\Big(1 - \frac{MN}{S}\Big)R^*,
 \end{split}
 \end{equation}
 where the factor ($S-S_{\mathrm{tr}}$)$/S$ accounts for the channel training overhead; the first inequality is due to $S_{\mathrm{tr}}\geq MN$; and the second inequality follows since ($\hat{\textbf{v}}[t],\hat{\textbf{f}}[t]$) is selected based on the estimated channel $\widehat{\bm{\Phi}}[t]$, which is usually different from the actual channel $\bm{\Phi}[t]$. The result in (6) shows that for IRS-aided communication when $MN$ is comparable or even exceeds $S$, which can easily occur as $M$ and $N$ are typically large (say $M=64$ and $N=10000$), there exists a huge gap between the practically achievable communication rate with the training-based beam selection taken into account and the maximum possible rate assuming perfect CSI.\par
Besides the prohibitive training overhead, another critical challenge for channel-training based beam selection lies in the high-computation overhead for beam selection in (5). For each channel coherent block, it may require exhaustive search over $|\mathcal{V}||\mathcal{F}|\gg 1$ beam pairs, which incurs prohibitive computational complexity as well. \par
 To address the above issues, we propose a novel environment-aware joint active and passive beam selection scheme for IRS-aided communication, based on the new concept of CKM \cite{17}. In particular, we propose environment-aware beam selection for IRS-aided communication enabled by one specific type of CKM, namely BIM \cite{18}, for which the active and passive beamforming vectors are selected based on the UE location, with either no real-time channel training required (i.e., training-free beam selection) or with only limited channel training (i.e., light-training beam selection), as detailed in the next section.

\section{Environment-Aware Beam Selection via Beam Index Map (BIM)}
Physically, wireless communication channels are mainly determined by the signal characteristics (such as wavelength), the transmitter/receiver locations, and the surrounding radio propagation environment. Conventionally, wireless channels are usually modeled statistically \cite{19}, based on probabilistic distributions on certain channel parameters (such as channel gain,  shadowing loss, existence of line-of-sight (LoS) path or not) and surrounding environment (such as building location and height distributions). Such channel modelling methods are quite useful for the performance analysis and design of wireless communication systems in an average sense, thanks to their mathematical tractability and generalizability.  However, they only use very coarse information of both the transmitter/receiver locations (such as their distance only rather than actual locations) and the propagation environment (such as urban, sub-urban or rural area only, while ignoring the actual environment where the communication takes place). As a consequence, the channel knowledge predicted by such models usually deviate significantly from the true channel values. Therefore, for each wireless communication pair in a given environment, pilot-based channel training is usually needed for CSI acquisition, which, however, suffers from high training overhead for IRS-aided communication. To address this issue, by exploiting the readily-available location information of UEs with ever-increasing accuracy in today's wireless networks, the environment-aware wireless communication enabled by CKM \cite{17} is practically appealing.\par
For the considered IRS-aided communication system, with the locations of the BS and the IRS fixed, the equivalent cascaded channel $\bm{\Phi}[t]$ varies mainly with the UE location, denoted as $\textbf{q}[t]\in\mathcal{Q}$, and the actual propagation environment, denoted as $E[t]$. Thus, $\bm{\Phi}[t]$  can be represented as
\begin{equation}
\bm{\Phi}[t] = g_{1}(\textbf{q}[t], E[t]), t=1,2,...,T,
\end{equation}
where $g_{1}(,.,)$ is an unknown function that is difficult to characterize exactly in practice. Fortunately, a novel approach to tackle this intractable problem is by utilizing CKM, which aims to provide a mapping between any potential UE location and its location-specific channel knowledge.\par
One useful example of CKM is \emph{channel matrix map} (CMM), which aims to directly predict the channel coefficient matrix $\bm{\Phi}[t]$ based on the UE location. However, as $\bm{\Phi}[t]\in \mathbb{C}^{N\times M}$ is of high dimension in our case, it will require massive storage capacity for CMM construction. Alternatively, it is observed from (4) and (7) that the optimal beam pair can be represented as an abstract function of the UE location and the propagation environment as
\begin{equation}
(\textbf{v}_{\mathrm{opt}}[t],\textbf{f}_{\mathrm{opt}}[t]) = g_{2}(\textbf{q}[t], E[t]), t=1,2,...,T,
\end{equation}
where $g_2(.,.)$ is a function that cascades $g_1(.,.)$ in (7) and the maximum operation in (4). Besides, in most practical scenarios, the major components of the radio propagation environment (such as the locations, height, and dielectric properties of the buildings) change in a much larger time scale than the UE location. As such, the variation of wireless channel is mainly due to user mobility. Therefore, for any given communication environment, (8) can be further simplified as
\begin{equation}
(\textbf{v}_{\mathrm{opt}}[t], \textbf{f}_{\mathrm{opt}}[t]) \approx g_{3}(\textbf{q}[t]), t = 1,2,...,T,
\end{equation}
where $g_{3}(,.,)$ is an abstract function that relates the beam pair with UE location for the given environment. Based on (9), a promising candidate of CKM, termed BIM\cite{18}, can be adopted for IRS-aided communication to select the beam pairs based on UE location in the given environment. Specifically, BIM establishes a mapping from the UE location in the specific area  $\mathcal{Q}$ to its best beam pair in the codebooks $\mathcal{V}$ and $\mathcal{F}$, i.e.,
\begin{equation}
\textbf{q}\in\mathcal{Q}\longrightarrow(\textbf{v}, \textbf{f})\in \mathcal{V,F}.
\end{equation}
\par
In practice, BIM can be constructed based on some finite data set that is obtained via various methods, such as dedicated offline data acquisition or online training. For offline data acquisition, ray-tracing based simulation or on-site measurement using automatic vehicles dispatched to the area of interest $\mathcal{Q}$ can be adopted. On the other hand, for online data acquisition, as the actual communication takes place, the location-specific beam pair obtained with any existing  channel estimation or beam training methods can be collected and stored in the database, rather than being discarded after use as in conventional communication systems. Let $I$ denote the total number of data entries available for BIM construction, and $\mathcal{Q}^{\prime}\subset\mathcal{Q}$ denote the corresponding UE locations where the preferred beam pairs are available based on offline or online data acquisition methods mentioned above. Thus, the data set for BIM construction can be expressed as the tuple $(\textbf{q}_{i}, \textbf{v}_{i}, \textbf{f}_{i}), i=1,...,I,$ where $\textbf{q}_{i}\in\mathcal{Q}^{\prime}$.\par
With the available data set for BIM construction,  we are  now ready to discuss two BIM-enabled beam selection schemes, which require no or light online beam training, respectively.
\subsection{BIM-Enabled Training-Free Beam Selection}
For the actual communication with known UE location $\textbf{q}[t]\in\mathcal{Q}$, which can be acquired via various contemporary  localization techniques with ever-increasing accuracy, such as GPS, Beidou, vision- or laser-based technologies, efficient beam selection can be achieved  based on the BIM. To this end, we first order the UE locations in the training dataset $\mathcal{Q}^{\prime}$ based on their distances to $\textbf{q}[t]$ with increasing order, which are denoted as $\textbf{q}_{[1]},...,\textbf{q}_{[I]}$, and their corresponding beam pairs are denoted as  $(\textbf{v}_{[1]}, \textbf{f}_{[1]}),...,(\textbf{v}_{[I]}, \textbf{f}_{[I]})$. Then for some pre-determined number $K\ll I$, the set $\mathcal{K}=\{(\textbf{v}_{[1]}, \textbf{f}_{[1]}), ...,(\textbf{v}_{[K]}, \textbf{f}_{[K]})\}$ form $K$ candidate beamforming pairs for the considered UE. Furthermore, denote the number of occurrence of distinct beamforming pairs in $\mathcal{K}$ as $w_{j}, j=1,2,...,J$, with $J\leq K$. Then the beam pair that appears most frequently in $\mathcal{K}$ will be selected for the considered UE at location $\textbf{q}[t]$. For the case when there is a tie, the one that gives the largest inverse-distance-sum from $\textbf{q}[t]$ will be selected. Fig. 2(a) illustrates the proposed BIM-enabled training-free beam selection scheme. Since BIM can be constructed offline and the online beam selection requires only the UE location, the effective communication rate for BIM-enabled training-free beam selection is
 \begin{equation}
   R = \frac{1}{T}\sum\limits_{t=1}^{T}\log_{2}(1 + \bar{P}|\hat{\textbf{v}}^{H}[t]\bm{\Phi}[t]\hat{\textbf{f}}[t]|^{2}),
 \end{equation}
 where ($\hat{\textbf{v}}[t],\hat{\textbf{f}}[t]$) is the selected beam pair at channel coherent block $t$ obtained based on the BIM and the UE location $\textbf{q}[t]$. By comparing (6) and (11), it is observed that the pre-log factor in (6) is avoided by the BIM-enabled training-free beam selection. Therefore, the achievable rate $R$ in (11) is expected to be closer to the optimal $R^{*}$ than (6).
\subsection{BIM-Enabled Light-Training Beam Selection}
As the BIM-enabled training-free scheme simply selects the beam pair among the $K$ nearest neighbors of the target UE based on their occurrence frequency, it is vulnerable to the variation of environment and measurement error. In order to improve the robustness, we further propose the BIM-enabled light-training beam selection scheme, which only requires low training overhead.\par
Similar to the training-free scheme, for the light-training beam selection scheme, the candidate set $\mathcal{K}$ for the consisdered UE location with $\textbf{q}[t]\in\mathcal{Q}$ is firstly formed based on its $K$ nearest neighbors. Instead of directly choosing the beam pairs with the highest occurrence frequency, each beam pair in the candidate set $\mathcal{K}$ is further trained online by beam sweeping, which thus yields,
\begin{equation}
   (\hat{\textbf{v}}[t],\hat{\textbf{f}}[t]) = \arg\mathop{\max}_{(\textbf{v},\textbf{f})\in\mathcal{K}}|\sqrt{P}\textbf{v}^{H}\bm{\Phi}[t]\textbf{f}x[t,s] + z[t,s]|^{2}.
 \end{equation}
Therefore, $K\ll MN$ beam sweeps are required and the effective communication rate is
 \begin{equation}
   R = \frac{1}{T}\sum\limits_{t=1}^{T}\frac{S-K}{S}\log_{2}(1 + \bar{P}|\hat{\textbf{v}}^{H}[t]\bm{\Phi}[t]\hat{\textbf{f}}[t]|^{2}).
 \end{equation}\par
Compared to (6), it is observed that the online training overhead $K$ for the BIM-enabled environment-aware beam selection is much smaller. Besides, thanks to the additional beam sweeping over the $K$ candidate beam pairs, the light-training scheme is expected to more likely select the optimal beam pair than the training-free scheme even under environment change and/or measurement error.
\begin{figure}
   \centering
   \subfigtopskip=2pt
   \subfigbottomskip=5pt
   \subfigcapskip=5pt
   \subfigure[BIM-enabled training-free]{
      \label{fig1}
      \includegraphics[scale=1.2]{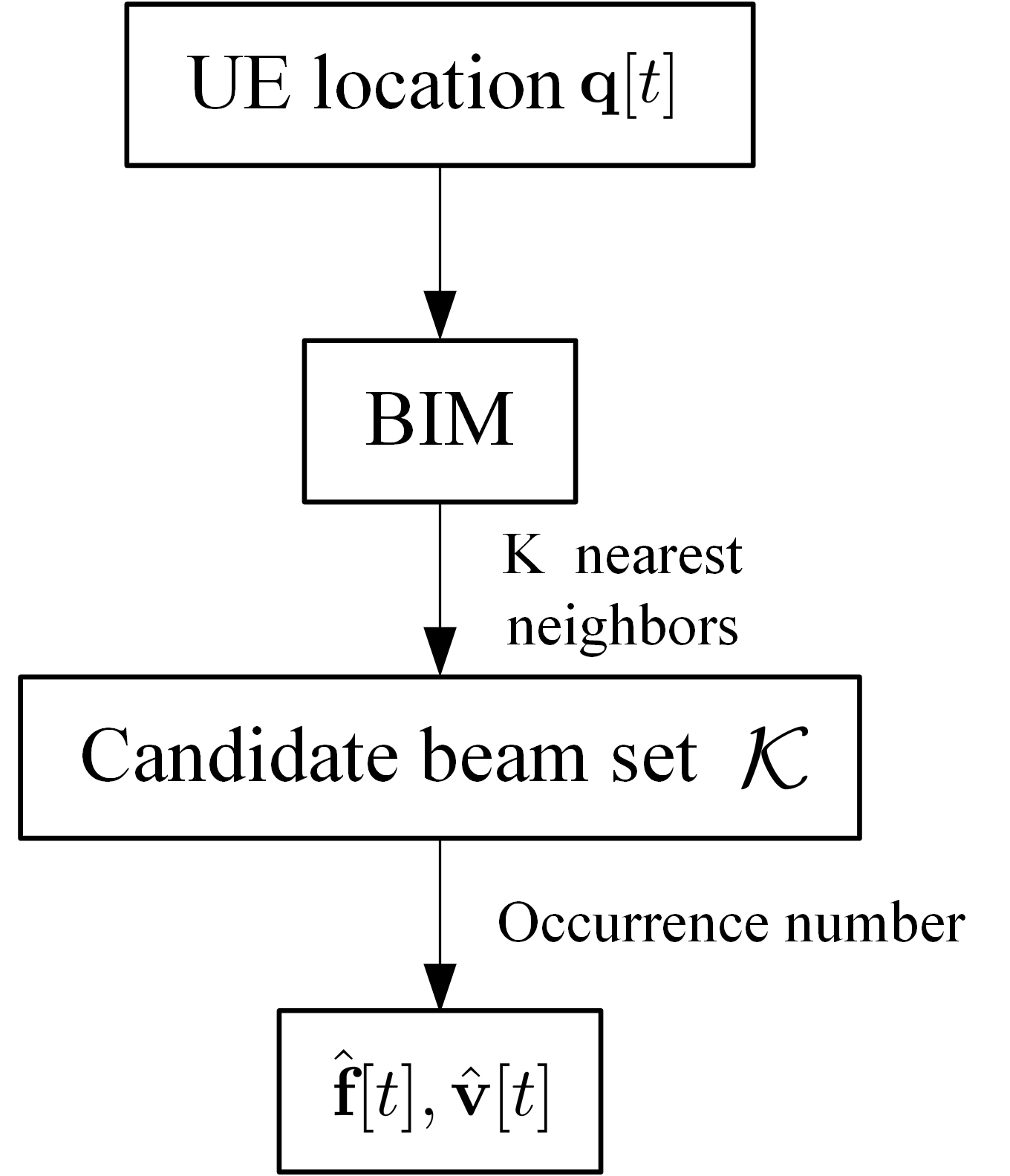}}
   \subfigure[BIM-enabled light-training]{
      \label{fig2}
      \includegraphics[scale=1.2]{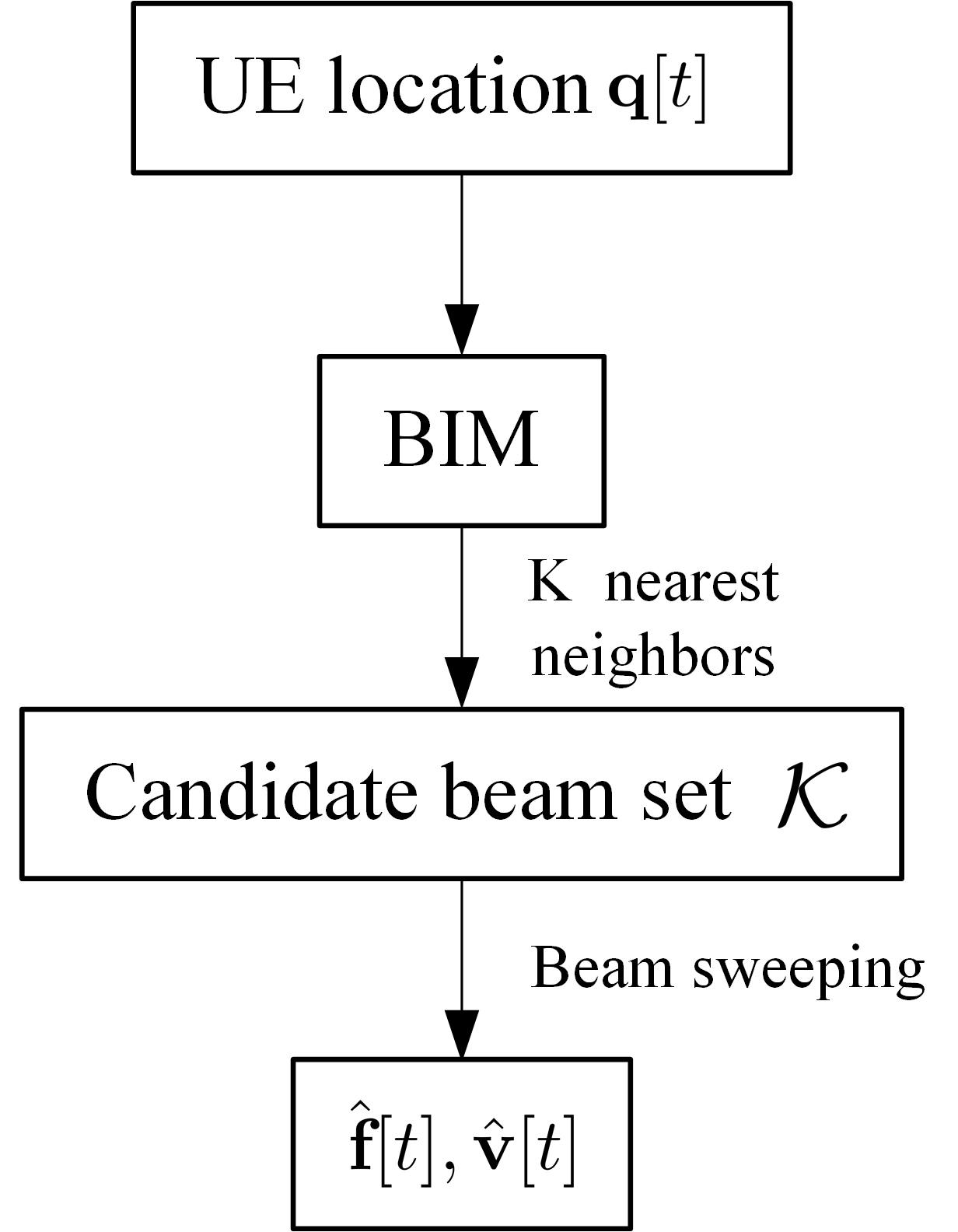}}
   \quad
   \subfigure[Location-based]{
      \label{fig3}
      \includegraphics[scale=1.2]{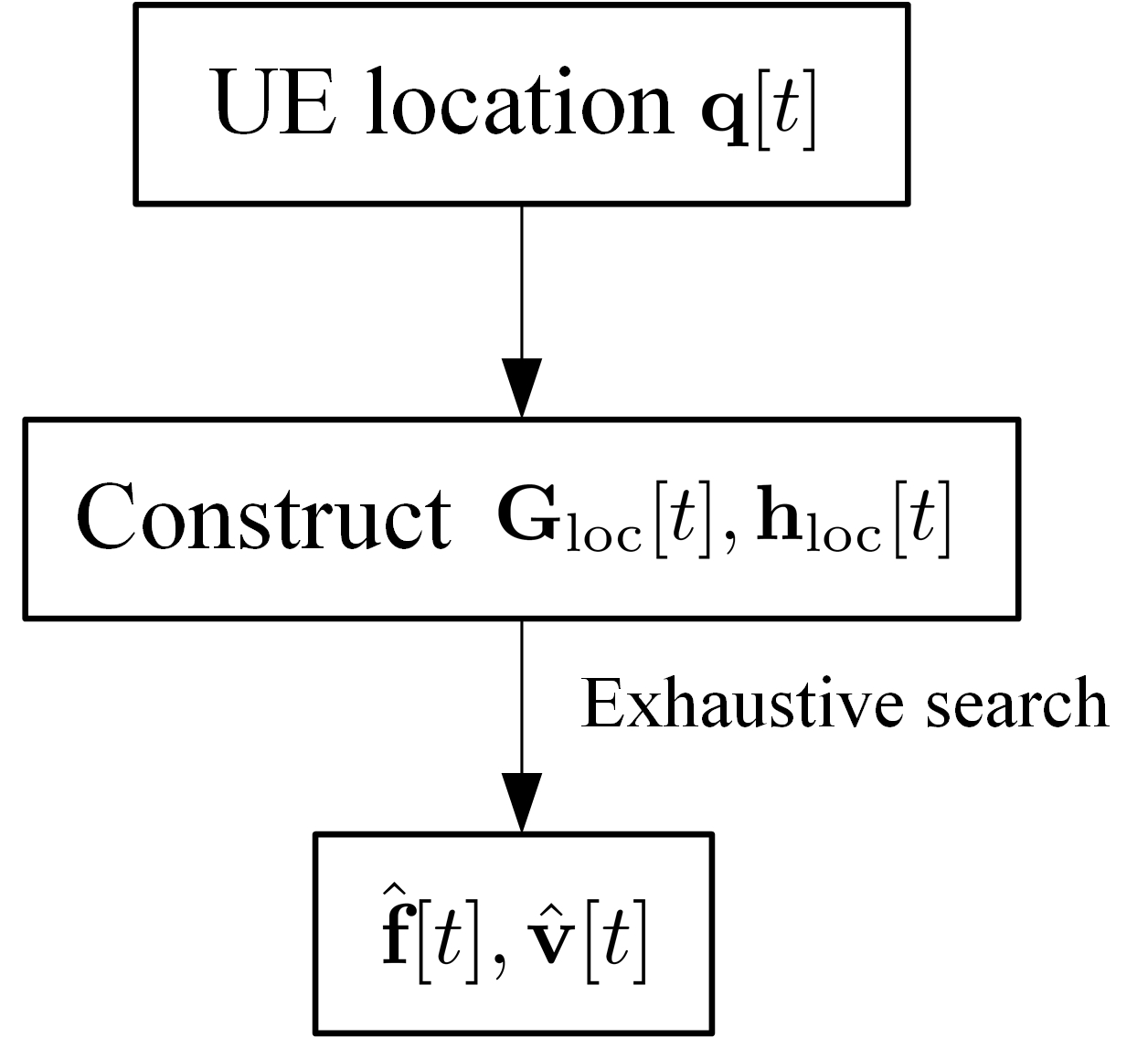}}
   \quad
   \subfigure[Two-time-scale based training]{
      \label{fig4}
      \includegraphics[scale=1.2]{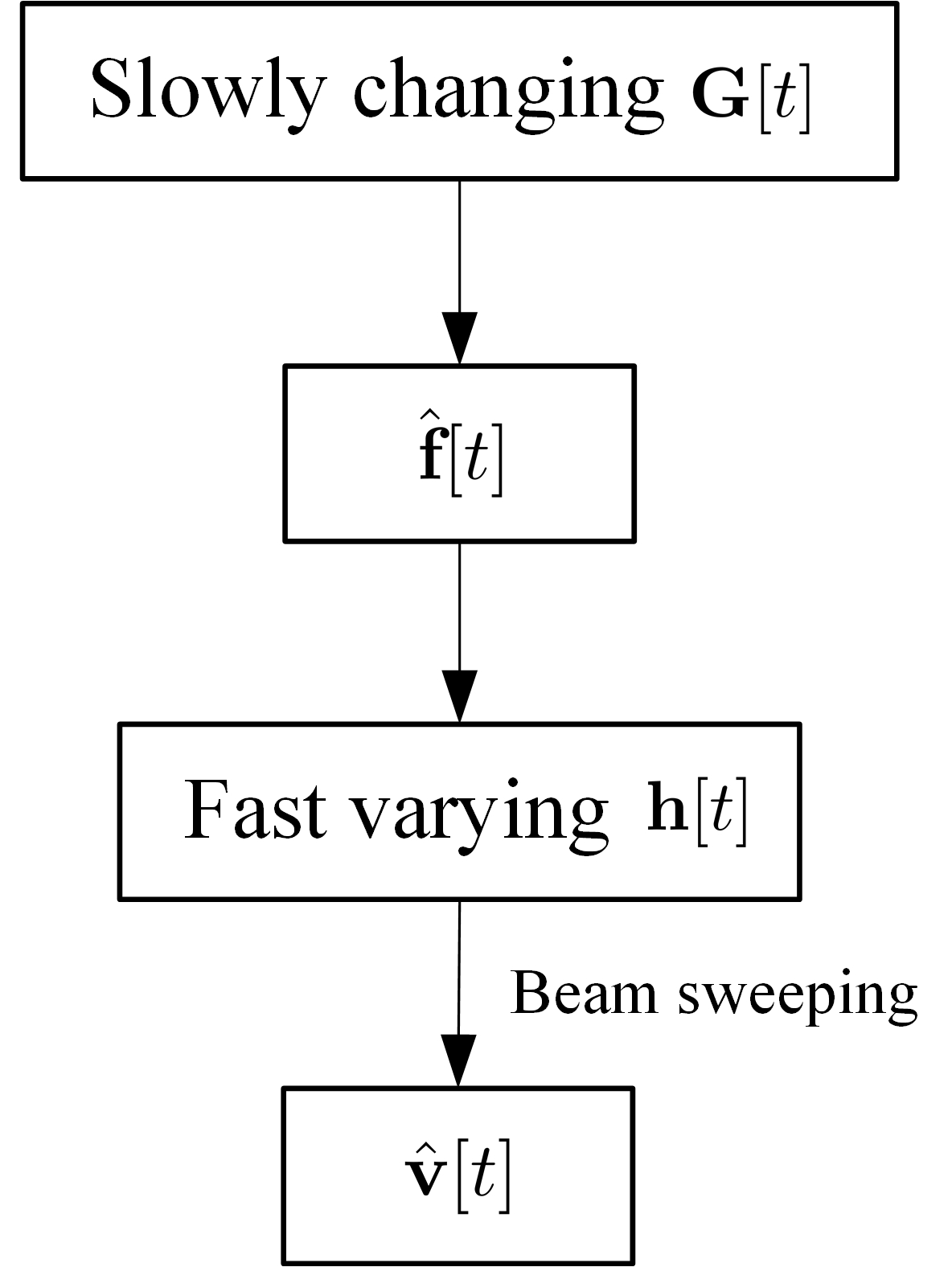}}
   \caption{An illustration of various beam selection schemes for IRS-aided communication.}
   \label{fig}
\end{figure}
\subsection{Benchmark Beam Selection Schemes}
In this subsection, we discuss two benchmark beam selection schemes that are unaware of the environment, namely the \emph{location-based beam selection} and the \emph{two-time-scale based beam sweeping}.\par
For the location-based beam selection scheme, the beam pair is selected based on the UE location only, while ignoring the actual propagation environment. In this case, the AoAs/AoDs are firstly calculated based on the locations of the BS, IRS, and UE. By further assuming the pure LoS link, the BS-IRS channel $\textbf{G}_{\mathrm{loc}}[t]$ and the IRS-UE channel $\textbf{h}_{\mathrm{loc}}[t]$ are approximately constructed as
\begin{equation}
\begin{split}
\textbf{G}_{\mathrm{loc}}[t]&=\alpha[t]\textbf{a}_{\mathrm{r}}(\theta_{\mathrm{AoA}}[t], \phi_{\mathrm{AoA}}[t])\textbf{a}_{\mathrm{t}}^{H}(\theta_{\mathrm{AoD}}[t], \phi_{\mathrm{AoD}}[t]),\\
\textbf{h}_{\mathrm{loc}}[t]&=\beta[t]\textbf{a}_{\mathrm{r}}(\theta^{\prime}_{\mathrm{AoD}}[t], \phi^{\prime}_{\mathrm{AoD}}[t]),
\end{split}
\end{equation}
where $\alpha[t]$ and $\beta[t]$ can be randomly chosen since they have no impact on the beam selection, $\theta_{\mathrm{AoA}}[t]$ and $\phi_{\mathrm{AoA}}[t]$ are the assumed zenith and azimuth AoAs solely based on the locations of the BS and IRS. Similarly for the AoDs $\theta_{\mathrm{AoD}}[t]$, $\phi_{\mathrm{AoD}}[t]$, $\theta^{\prime}_{\mathrm{AoD}}[t]$ and $\phi^{\prime}_{\mathrm{AoD}}[t]$, $\textbf{a}_{\mathrm{r}}(,.,)$ and $\textbf{a}_{\mathrm{t}}(,.,)$ respectively represent the array response vector associated with the IRS and the BS. It then follows from (4) and (14) that the objective function for location-based beam selection can be expressed as
\begin{equation}
\begin{split}
&|\textbf{v}^{H}\mathrm{diag}(\textbf{h}_{\mathrm{loc}}^{H}[t])\textbf{G}_{\mathrm{loc}}[t]\textbf{f}|^{2}\\
=&|\textbf{v}^{H}\mathrm{diag}(\textbf{h}_{\mathrm{loc}}^{H}[t])\alpha[t]\textbf{a}_{\mathrm{r}}(\theta_{\mathrm{AoA}}[t], \phi_{\mathrm{AoA}}[t])\\
&\cdot\textbf{a}_{\mathrm{t}}^{H}(\theta_{\mathrm{AoD}}[t], \phi_{\mathrm{AoD}}[t])\textbf{f}|^{2}\\
=&|\textbf{v}^{H}\mathrm{diag}(\textbf{h}_{\mathrm{loc}}^{H}[t])\alpha[t]\textbf{a}_{\mathrm{r}}(\theta_{\mathrm{AoA}}[t], \phi_{\mathrm{AoA}}[t])|^{2}\\
&\cdot|\textbf{a}_{\mathrm{t}}^{H}(\theta_{\mathrm{AoD}}[t], \phi_{\mathrm{AoD}}[t])\textbf{f}|^{2}.
\end{split}
\end{equation}
It is observed from (15) that maximizing the beamforming gain can be achieved by selecting the optimal $\textbf{v}[t]$ and $\textbf{f}[t]$ independently, i.e.,
\begin{equation}
\begin{split}
&\hat{\textbf{v}}[t] = \arg\mathop{\max}_{\textbf{v}\in\mathcal{V}}|\textbf{v}^{H}\mathrm{diag}(\textbf{h}_{\mathrm{loc}}^{H}[t])\textbf{a}_{\mathrm{r}}(\theta_{\mathrm{AoA}}[t], \phi_{\mathrm{AoA}}[t])|^{2},\\
&\hat{\textbf{f}}[t] = \arg\mathop{\max}_{\textbf{f}\in\mathcal{F}}|\textbf{a}_{\mathrm{t}}^{H}(\theta_{\mathrm{AoD}}[t], \phi_{\mathrm{AoD}}[t])\textbf{f}|^{2}.
\end{split}
\end{equation}
Although the location-based beam selection scheme only requires the location information, its ignorance of the actual radio propagation environment may result in large discrepancy between the selected beam pair and the optimal one when the LoS link is blocked and/or when there exists strong non-line-of-sight (NLoS) paths.\par
On the other hand, by assuming that the BS-IRS channel varies much more slowly compared to the IRS-UE channel, the channel estimation overhead in (6) can be significantly reduced by a two-time-scale beam  sweeping scheme. Specifically, the BS-IRS channel $\textbf{G}[t]$ is assumed to remain approximately constant over the $T$ blocks, which is assumed to be  known and denoted as $\textbf{G}$. Based on $\textbf{G}$, the transmit beamforming vector at the BS is firstly selected to maximize the total power towards the IRS (e.g., the IRS controller), i.e.,
\begin{equation}
   \hat{\textbf{f}} = \arg\mathop{\max}_{\textbf{f}\in\mathcal{F}}\Vert{\textbf{G}\textbf{f}}\Vert^{2}.
\end{equation}
Based on the selected active beam $\hat{\textbf{f}}$, the passive beamforming vector $\hat{\textbf{v}}$ is then selected online by an exhaustive beam sweeping over all beams in the codebook $\mathcal{V}$, which requires training overhead $|\mathcal{V}|$.
Therefore, the effective average communication rate is expressed as
\begin{equation}
   R = \frac{1}{T}\sum\limits_{t=1}^{T}\frac{S-|\mathcal{V}|}{S}\log_{2}(1 + \bar{P}|{\hat{\textbf{v}}[t]}^{H}\bm{\Phi}[t]{\hat{\textbf{f}}}|^{2}).
\end{equation}
By comparing (6) and (18), it is observed that the training overhead with the two-time-scale beam sweeping scheme is much smaller than that for the conventional channel training-based scheme. However, the training overhead is still significant since $|V|\gg 1$. Besides, as the selection of $\hat{\textbf{f}}$ in (17) ignores the IRS-UE channel, the selected beam pair may deviate from the optimal one.\par
The procedures of location-based and two-time-scale based beam selection are illustrated in Fig. 2(c) and (d), respectively.

\section{Simulation Results}
As shown in Fig. 3, we consider an actual physical environment with two large buildings and many small scatterers. The locations of the BS and IRS are fixed, while the UE is randomly distributed in a 10m$\times$10m square area near the IRS. The BS and IRS are placed such that there exists an LoS path between them, while the direct BS-UE channel is blocked severely by an intermediate building. There are also other blockers/scatterers randomly located in the considered square area (shown in Fig. 3), which may block the LoS link and/or create multi-paths for the IRS-UE channel. The commercial ray tracing software Remcom Wireless Insite\footnote{https://www.remcom.com/wireless-insite-em-propagation-software} was used to generate the actual CSI, including the power, phase, AoA, AoD of all channel paths at each location.\par
\begin{figure}[h]
\centering
\includegraphics[width=7.5cm,height=5cm]{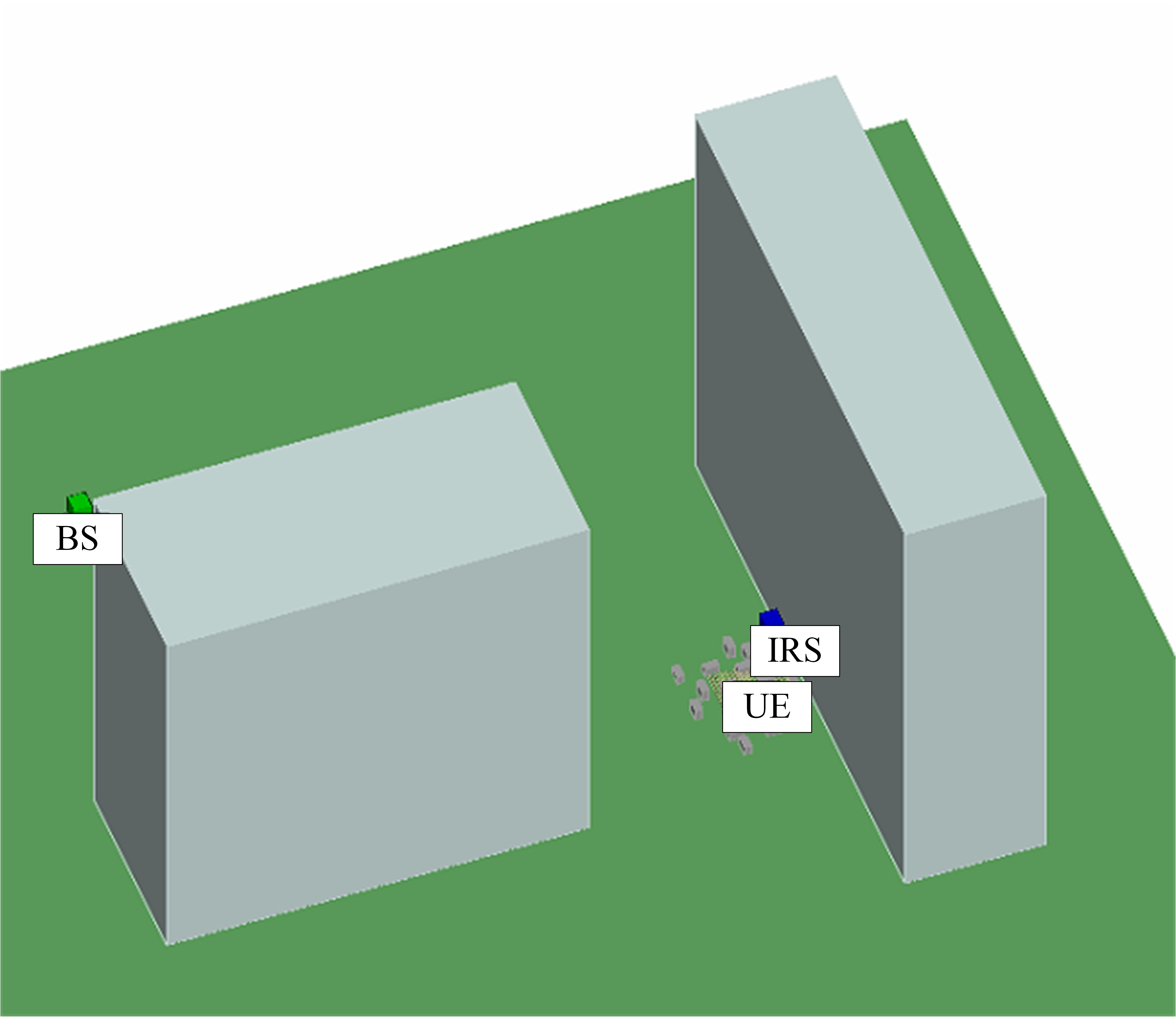}
\caption{Physical environment for numerical simulations.}
\end{figure}
The BS is assumed to be equipped with an $8\times8$ uniform planar array (UPA) with adjacent elements separated by half wavelength, and a Kronecker product based beamforming codebook is used at the BS. Besides, the IRS is composed of $100\times100$ reflecting elements with a DFT based passive beamforming codebook \cite{10}. The noise power spectral density and the system bandwidth are set as -174 dBm/Hz and 10 MHz, respectively. The transmit power $P$ at the BS varies from 10 dBm to 40 dBm.\par
A BIM is constructed to learn the best beam pairs based on the training data set composed of 984 randomly selected UE locations in the considered square area. Based on the ray tracing data, the actual channel matrices are constructed and the best beam pair for each UE location in the training set is obtained based on (4). The number of nearest neighbors $K$ used for beam selection is 3. Besides, each channel coherent block is assumed to span $S = 2\times10^{4}$ symbols, and all the average communication rate values obtained are based on 100 randomly selected UE locations.\par
\begin{figure}[h]
\vspace{-0.2cm}
\centering
\includegraphics[width=8.7cm,height=6.5cm]{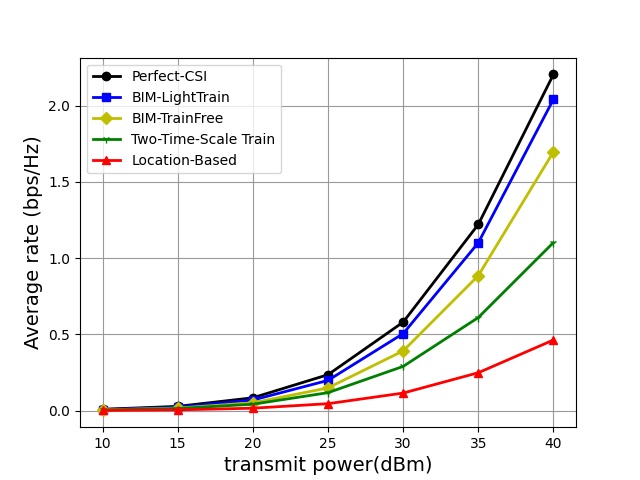}
\caption{Comparison of various beam selection schemes with perfect UE location.}
\end{figure}
Fig. 4 shows the average effective communication rate versus the BS transmit power for various beam selection schemes. The UE location, which is required for both the location-based benchmark scheme and the proposed BIM-based scheme, is assumed to be perfectly known in this example. It is observed that both the proposed BIM-enabled training-free and light-training beam selection schemes significantly outperform the benchmark location-based and two-time-scale based schemes, thanks to their great reduction in training overhead and environment awareness. Besides, though requiring moderate additional training overhead, light-training scheme further outperforms the training-free scheme, and its performance is quite close to that of the ideal case with perfect CSI. It is also observed that the two-time-scale based beam sweeping scheme has relatively poor performance since it still requires the high training overhead for selecting the passive beamforming vector. Moreover, the location-based scheme gives the worst performance, since it ignores the actual environment with blockers and scatters present.\par
\begin{figure}[h]
\vspace{-0.2cm}
\centering
\includegraphics[width=8.7cm,height=6.5cm]{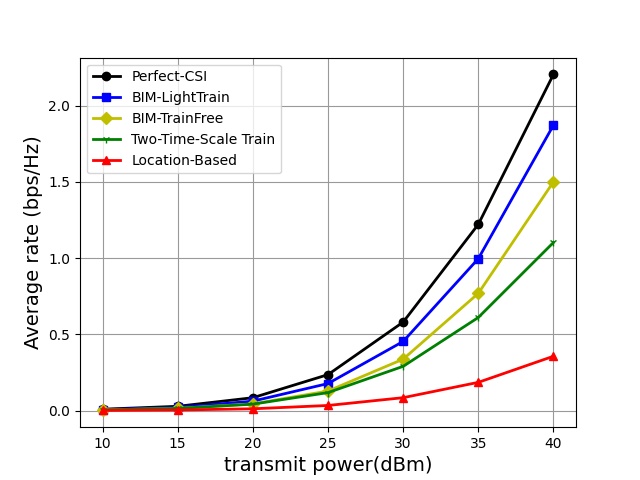}
\caption{Comparison of various beam selection schemes with an average UE location error of 0.3 meter.}
\end{figure}
In order to study the impact of UE location error, Fig. 5 shows the average communication rate when the UE location is subjected to random errors following Rayleigh distribution, with an average error of 0.3 meter. By comparing Fig. 4 and Fig. 5, it is observed that the rate performance of both the location-based and BIM-based schemes are slightly degraded. However, even with such moderate location error, the proposed BIM-based schemes still significantly outperform the benchmark location-based and two-time-scale based schemes. This demonstrates their robutness to UE location errors in practice for IRS-aided communication.
\addtolength{\topmargin}{0.08in}
\section{Conclusion}
In this paper, we proposed environment-aware beam selection schemes for IRS-aided communication enabled by the BIM, which requires no online training or only limited beam training. Based on the realistic communication channels obtained by ray-tracing, simulation results demonstrated that the proposed schemes yield significant rate improvement over the benchmark schemes without using the BIM, and are also quite robust to UE location errors in practice.

\section*{Acknowledgment}
This work was supported by the National Key R\&D Program of China with Grant number 2019YFB1803400.

\bibliographystyle{IEEEtran}

\end{document}